# Effect of dopants' concentration on high-dose high-temperature thermoluminescence of LiF:Mg,Cu,P detectors: Mg and Cu influence

Barbara Obryk[a*], Mariusz Kłosowski[a], Patrycja Baran[b,d], Krzysztof Hodyr[c]

[a]Institute of Nuclear Physics Polish Academy of Sciences (IFJ), ul. Radzikowskiego 152, 31-342 Kraków, Poland

[b]AGH University of Science and Technology, Al. Mickiewicza 30, 30-059 Kraków, Poland

[c]Institute of Applied Radiation Chemistry, Lodz University of Technology, ul. Wróblewskiego 15, 90-924 Łódź, Poland

[d]ARC Centre of Excellence in Advanced Molecular Imaging, School of Physics, The University of Melbourne, Parkville 3010, Australia

**HIGHLIGHTS**

- **Variation of dopants' concentration is influencing the shape of glow-curve.**
- **Magnesium presence is crucial for occurrence of the TL peak 'B'.**
- **Mg influences the shape of the glow curve and high-temperature peaks intensity.**
- **Cu influences high-temperature peaks intensity mainly.**

**Abstract**

In search of origin and mechanism of the high-dose high-temperature TL emission with presence of the peak 'B' about thirty batches of LiF:Mg,Cu,P detectors with different dopants' concentration were produced and comprehensive studies of their high-dose TL features have been performed. Within each step of the research, these characteristics were compared with the high-dose TL properties of the typical MCP-N detectors' batch highly exposed simultaneously with the detectors from batches of varying dopants concentration. We determined the effect of concentration of specific dopants on the TL peak 'B'. It was found that both magnesium and copper are necessary to obtain high-dose TL signal, however, the important for high-dose TL features seems to be also concentration of phosphorus. In the framework of this study we present results of magnesium and copper concentration changes to the TL signal after high-doses of electron radiation.

* Corresponding author: e-mail: barbara.obryk@ifj.edu.pl; tel.: +48-12-6628280; fax: +48-12-6628066

# 1. Introduction

The high-dose properties of the LiF:Mg,Cu,P phosphor, observed at the IFJ in 2006, manifesting as high temperature TL emission with maximum at about 400°C (denoted as the TL peak 'B') [Bilski et al., 2008, 2010; Obryk et al., 2009; 2010; 2011a], allow for the dose calibration with this phosphor in the kilogray range [Obryk et al., 2011b; Obryk, 2013]. Since LiF:Mg,Ti phosphor high-dose features are very limited [Bilski et al., 2010, Khoury et al., 2011], therefore it is believed that the dopants play the lead role in this phenomenon [Obryk, et.al, 2014, Remy et al., 2016]. Until now, studies of dopants concentration influence on TL emission were conducted in detail for the LiF based phosphors in the mGy dose range, where TL glow-curve is a typical one without the high-temperature TL peak 'B' presence [e.g. Bos et al., 1996; Bilski et al., 1996; 1997; Chen and Stoebe, 1998; 2002; Lee et al., 2008; Patil and Moharil, 1995; Shoushan, 1988; Tang et al., 2008].

In a typical LiF:Mg,Cu,P detector [Nakajima et al., 1978], there are three types of dopant: magnesium, copper and phosphorous. The most important dopant responsible for the formation of trapping centers is magnesium. It was observed that even small changes in the concentration of magnesium produce a significant change in the shape of the glow-curve and the intensity of the individual peaks for low dose range [e.g. Bilski et al., 1997]. In the case of zero magnesium concentration practically only peak 4 (with maximum intensity at about 220°C) is observed. It is also noted that the area under the peak at the right side of the glow-curve increases with increasing magnesium concentration reaching a maximum for a certain concentration of magnesium. However, the value of this maximum is related to the concentration of copper in the detector. In addition, it was observed that the proportion of the low-dose higher temperature peaks to the total TL signal rises nearly in proportion to the concentration of magnesium in the detector [Bilski et al., 1996].

The role of copper as activator, despite a significant effect on the TL signal, even at low doses is not entirely clear. The concentration of copper has a strong influence on all the peaks of the glow-curve, both peak 4 and higher temperature peaks for low doses. It was observed that the response of the TL peak 4 increases with the increase of copper concentration to a value of 0.02 mol%, and then decreases at higher concentrations [Bos et al., 1996]. For detectors with zero copper content it appears that the higher temperature peaks reached the maximum, and increase in concentration reduces the amount of these peaks for low doses. It is also noted that the same level of higher temperature peaks are obtained with much lower concentration of copper, when the content of magnesium is reduced accordingly. Therefore, the effect of copper appears to be dependent on the magnesium content, copper can inhibit the effect of magnesium excess in the detector [Bilski et al., 1997].

Phosphorus is necessary to achieve high intensities of the peak 4 for low doses. This means that the intensity of the peak 4 depends on both the magnesium and phosphorus. With increasing concentration of phosphorus the amplitude of the signal increases to a threshold value. Above this threshold there is no correlation between the phosphorus content and the intensity of the peak 4 at low doses [Bilski1996].

All mentioned above concern low-dose range studies where the high-temperature peak 'B' is not present. A systematic investigation of the high-dose (kGy range) TL properties of LiF:Mg,Cu,P samples with different concentration of all activators has been performed at the IFJ recently. The effect of varying content of one of three dopants sequentially, namely magnesium, copper and phosphorus, while maintaining the typical concentration of the other two dopants in each case, have been studied. About thirty batches of sintered chips with dopants varied over the following ranges: 0-0.4 mol% Mg, 0-0.1 mol% Cu, 0-1.875 mol% P, were produced and comprehensive studies of their high-dose TL features in the different radiation fields (Co-60 gammas, 6 MeV electrons, 23 GeV protons, and reactor neutrons) have been performed. In this way, the impact of the type of dopants and their concentration on the occurrence and characteristics of the high-dose high-temperature TL peak 'B' has been verified. Comprehensive results of these studies will be presented in upcoming articles. In this work, being the first one of a series planned, we present results of magnesium and copper concentration changes to the TL signal after high-doses of electron radiation.

## 2. Materials and methods

### 2.1. Materials and synthesis

All detectors' batches have been produced at the IFJ in Kraków using sintering method. First lithium fluoride was synthesized in chemical processes between LiCl (AlfaAesar 99% reagent grade) and HF >48% (Sigma-Aldrich, puriss p. a.). Obtained LiF powder was thoroughly mixed with activators of the following forms: $MgCl_2$ solution (obtained from magnesium oxide POCH S.A. pure p.a. with fuming hydrochloric acid for trace analysis by Sigma-Aldrich), $CuCl_2$ solution (obtained from cooper oxide POCH S.A. pure p.a. with fuming hydrochloric acid for trace analysis by Sigma-Aldrich) and $H_3PO_4$ (POCH S.A> pure p.a.). The dopants were added in concentration related to LiF listed in Table 1. The pellets of 4.5 mm diameter and 0.9 mm thickness have been formed mechanically from each powder type. The last stage was sintering of pellets at a temperature between 600-700ºC in gas atmosphere and platinum containers for a certain period of time, as due to this stage LiF:Mg,Cu,P phosphor gains higher sensitivity. Two step pre-irradiation annealing of all samples has been applied as routinely for LiF:Mg,Cu,P detectors, i.e. 260ºC for 10 minutes followed by immediate cooling and then 240ºC for 10 minutes finalized with immediate cooling also.

**Table 1. The concentration of individual dopants for the studied detectors' batches.**

| Sample's mark | Dopants' concentration | | |
|---|---|---|---|
| | $Mg^{2+}$ [%mol] | $Cu^{2+}$ [%mol] | $PO_4^{3-}$ [%mol] |
| Ref | 0,2 | 0,05 | 0,625 |
| 0M | 0 | 0,05 | 0,625 |
| 0.5M | 0,1 | 0,05 | 0,625 |
| 2M | 0,4 | 0,05 | 0,625 |
| 0C | 0,2 | 0 | 0,625 |
| 0.5C | 0,2 | 0,025 | 0,625 |
| 2C | 0,2 | 0,1 | 0,625 |

### 2.2. Irradiation and measurements

Detectors were exposed using an electron linear accelerator (Linac) at the Laboratory of Linear Electron Accelerator at the Institute of Applied Radiation Chemistry of the Lodz University of Technology, Lodz, Poland. Detectors have been exposed with 6 MeV electron beam with dose rate of 6.1 kGy/min. The distance from the output window of the accelerator was 160 cm. In order to analyse the high-dose changes in the glow-curve for different types of detectors ten steps of doses have been selected (i.e. 1, 5, 10, 20, 30, 50, 100, 200, 500, 1000 kGy) and applied. Dosimetry was performed during the irradiation by the radiochromic foils.

The readout were made using Harshaw 3500 manual TL reader. The reading conditions for all detectors were the same: temperature range from 100°C to 550°C, a linear heating rate of 2°C/s, high-dose filter applied (CVI Melles Griot Laser Optics) which task was to reduce the signal reaching the photomultiplier in order to avoid its' malfunction due to not sufficient dynamic of its response.

## 3. Results and discussion

For each production batch the low-dose sensitivity has been evaluated. For this purpose several detectors from each production batch have been selected and irradiated with a dose of 1 mGy, using Cs-137 gamma source. In addition, reference LiF:Mg,Cu,P detectors were exposed simultaneously. On the basis of average TL signal values for detectors of each batch and for the reference detectors the sensitivity was determined as the ratio of these two values. From the viewpoint of low-doses the batch with sensitivity closest to the reference detectors is the one with copper concentration reduced 50% to the standard one. Other batches are not showing sufficient low dose sensitivity in comparison to the reference MCP-N phosphor, as its value vary from 37.6% for 0C batch to 71.0% for 2C (only 0M batch low dose sensitivity is not acceptable being 0.2% only).

The impact of the type of dopants and their concentration on the occurrence and characteristics of the high-dose TL signal has been verified. Figure 1 presents the TL glow-curves for the batches of detectors listed in Table 1 after exposures to selected doses (i.e. 1, 10, 100, 1000 kGy). At a dose of 1 kGy minimum intensity in relation to the reference detectors' glow-curve exhibit 0C, 0M and 0.5C batches. TL peak intensity for the other detectors is slightly higher than for the reference one. The shape of glow-curve for all groups of detectors (except 0M) is very similar. The only difference is the lack of peak 2 for all detectors with an altered concentration of dopants. Larger changes in the shape of the glow-curves are present for detectors exposed to a dose of 5 kGy, but also in this case the smallest TL signal is observed for the batches 0C, 0M and 0.5M. The highest intensity peaks were obtained for the detectors batches 2C and 2M.

The glow-curves obtained for intermediate doses – tens of kGy show a wide variation of shape. The weakest TL signal demonstrate detectors with zero copper content. For doses of 10 kGy and 20 kGy similarity can be seen in glow-curve shape between the groups of reference and 2C detectors, and between 2M and 0.5C also. The differences begin to be visible for the dose of 30 kGy and 50 kGy, while in the case of the first pair of detectors more dynamic change in shape of the glow-curve can be seen for reference detectors, which results in the earlier formation of the peak 'B'. Similar behaviour can also be seen in the case of the second pair, where the peak 'B' is present 'earlier' than for the 0.5C batch. For a dose of 50 kGy most similar to each other in shape are glow-curves of reference and 0.5C detectors.

From the glow-curves obtained for the third dose range (100-1000 kGy) it can be seen that they exhibit less variation in shape than in case of tens kGy. The highest change in the glow-curves of 100 kGy and 200 kGy is observed for the detectors from the group 0.5M, for which in case of higher doses the highest amplitude of the peak 'B' is shown. The glow-curves of the detectors 0C, 0.5C and 2M exhibit lower amplitude of the peak 'B' than the reference detectors. Satisfactory results in terms of intensity were obtained for the detectors 2C and 0.5M.

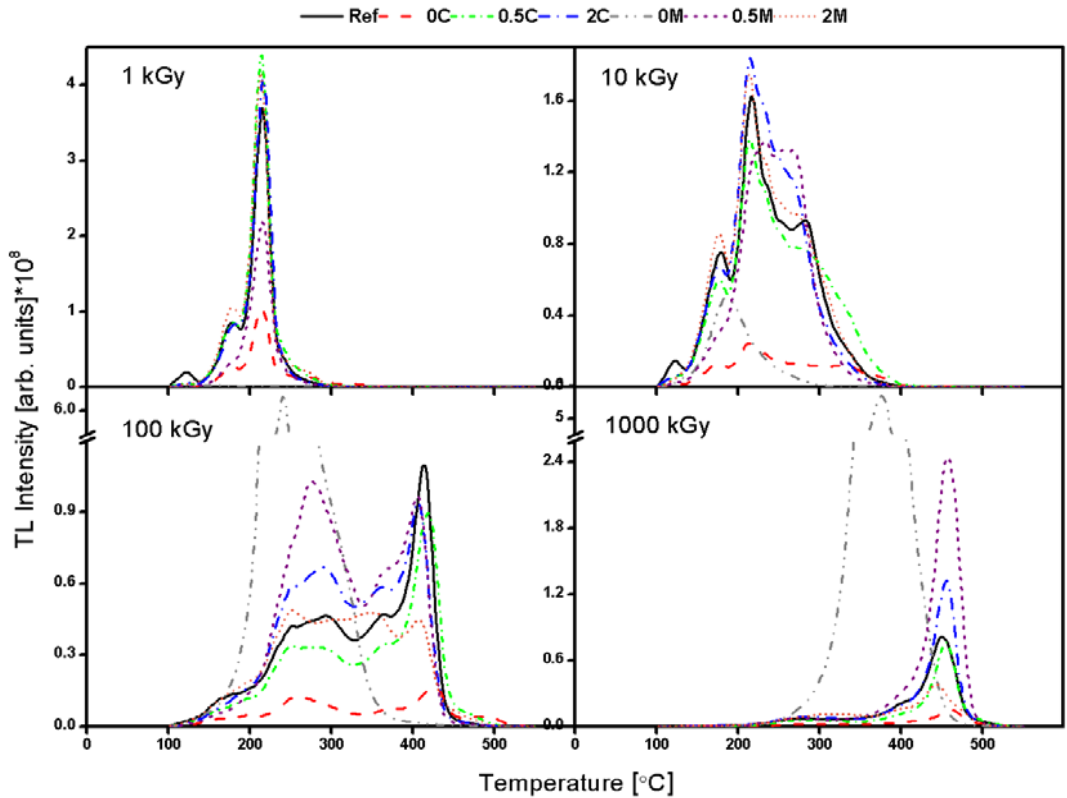

Figure 1. Glow curves' shape of LiF:Mg,Cu,P detectors with different copper and magnesium concentration exposed with: 1, 10, 100 kGy, and 1000 kGy of electrons, respectively.

Analysing the shape of the glow-curves it can be concluded that magnesium is necessary to the formation of high-dose high temperature peaks. After the exposure dose from the first range (1-5 kGy) it is observed that high-temperature peaks move towards higher temperature. A similar pattern can be seen in the case of the intermediate doses (10-50 kGy), while for lower concentration of magnesium the shift towards lower temperature as compared to the reference glow-curve is visible. It is interesting that for the highest dose a shift of the peak 'B' dependence on the magnesium concentration is not observed. The only differences are visible in the peak intensity which is highest for a reduced content of Mg. According to the previous studies about the effect of the concentration of copper at low doses (mGy range), beyond a certain concentration value the drop the TL signal is visible. It turned out, however, that for the high-dose TL signal increasing the concentration of copper in the detector allows to obtain a higher TL signal. Furthermore, detectors with a reduced content of Cu give much lower TL signal than the reference detectors, which also does not coincide with the low dose studies.

Figure 2 presents the values of the average total TL signal as a function of dose for all researched detectors' batches. The weakest TL response for doses up to 10 kGy feature detectors 0C and 0M. However, for 0M detectors signal start to grow for higher doses, while for 0C not. 0M batch total TL signal dominates for doses starting from 50 kGy. Generally total TL signal grows with growing Cu concentration for higher doses while for Mg concentration this is not the case. In the range of hundreds kilogray lower concentration of Mg than reference one results in higher total TL signal.
5

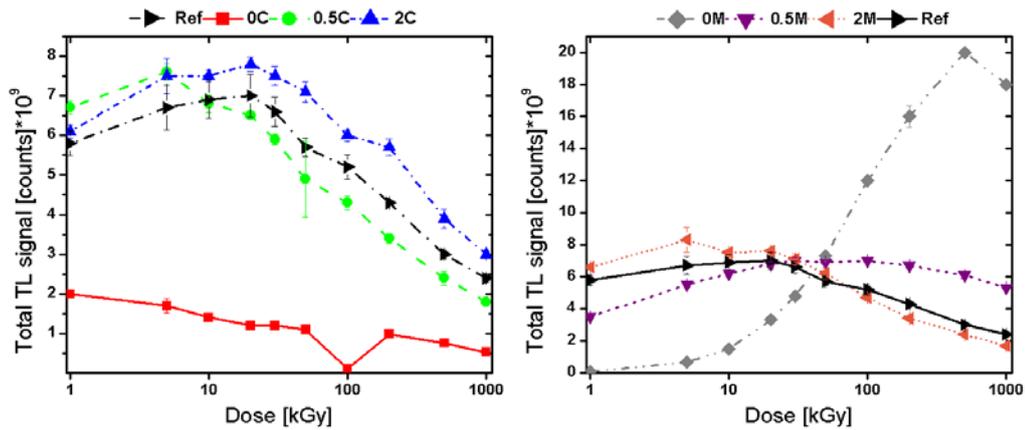

Figure 2. The average total TL signal (integrated in the range 100-550ºC) for all types of detectors exposed to 1-1000 kGy of electrons.

Figure 3 shows the share of the TL signal above 250ºC (UHTR(250) coefficient) and 350ºC (UHTR(350) coefficient) in the total TL for all detectors' types as a function of dose [Obryk et al., 2011b]. Analysing the results it can be noted that all curves except for the group of 0C detectors have similar shape. The results allow to conclude that the UHTR(250) can be used in dosimetry for doses ranging from 1 kGy to 100 kGy, where it is saturated. Noticeable differences were obtained for the UHTR(350) coefficient. However, the results obtained for the 2C detectors' type are similar to the reference one, which means that in this case application of an UHTR(350) allows to measure the dose from about 20 kGy to 1000 kGy. The best UHTR(350) curve seems to be obtained for 0.5M detectors' type, for which saturation of the TL signal is not observed, and the lower limit of the application of UHTR(350) seems to be about 50 kGy.

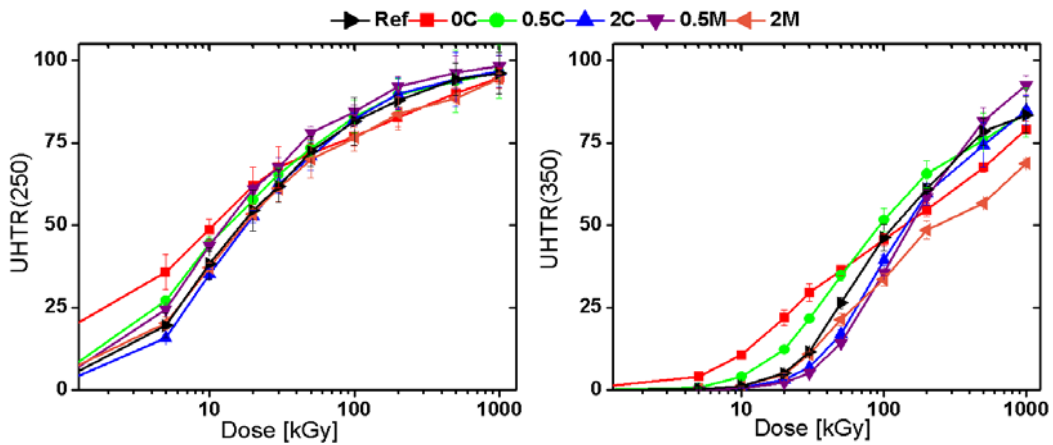

Figure 3. The dependence of the UHTR(250) and UHTR(350) coefficients as a function of the dose for all types of detectors.

An important issue in terms of high-dose dosimetry, i.e. improvement of the UHTR method [Obryk et al., 2011b], is the change the peak 'B' position depending on the dose as its' shift with the temperature could be also used to determine the dose (see Figure 4). This discussion does not include detectors 0M, since peak 'B' does not exist for them. For reference detectors, as well as 0C and 0.5C peak 'B' is already visible for a dose of 50 kGy, whereas for detectors 2C, 2M and 0.5M it appears at a dose of 100 kGy only. It was observed that in any case the temperature of peak 'B' increases with dose. The highest difference in the peak 'B' position for doses from 50 kGy to 1000 kGy was observed for the reference detectors, it amounts 57°C ± 2°C. For detectors 0C and 0.5C difference is equal and amounts to 48°C ± 2°C. For other types of detectors, the highest difference in peak 'B' position from the dose of 100 kGy to 1000 kGy was achieved with 0.5M (52°C ± 1°C) and the lowest for 2M (36°C ± 1°C).

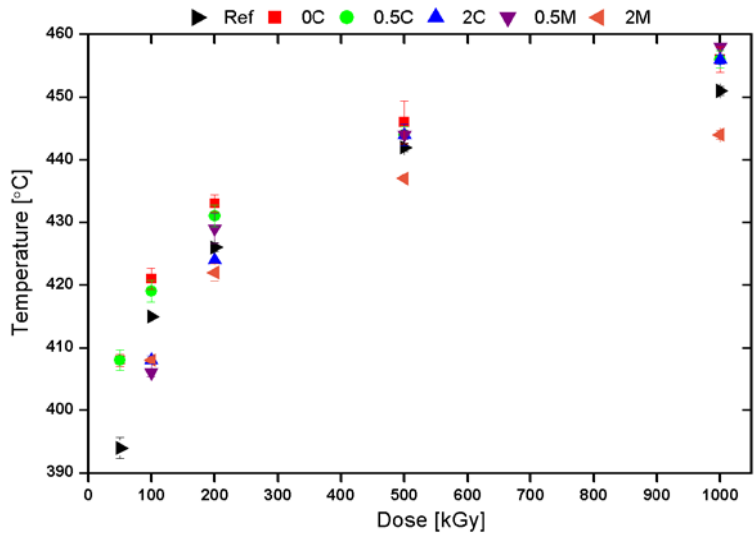

Figure 4. Comparison of the high-temperature TL peak 'B' position as a function of the dose for all types of detectors.

The maximum peak intensity is not for each glow-curve for which the peak 'B' is visible, the intensity of this peak. Therefore, the analysis was carried out, which compares the intensity of the peak 'B' obtained after exposure to the same dose in reference to the one obtained with the reference detector (see Figure 5). The highest intensity of the peak 'B' was obtained for 0.5M detectors, but it appears only at dose of 100 kGy. In the case of detectors with an altered concentration of the copper best values were obtained for the detectors 2C, where the peak is well visible already for 50 kGy. Other types of detectors in most cases are characterized by a reduction of the peak 'B' intensity with respect to the reference detectors.

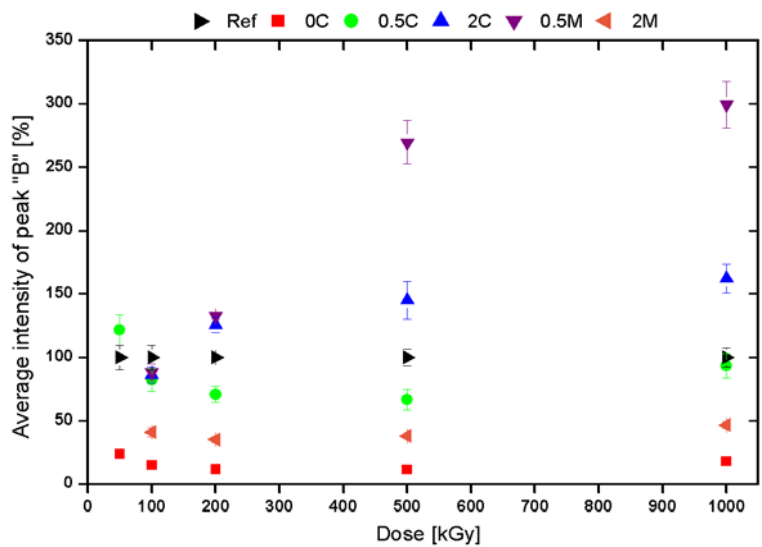

Figure 5. Comparison of the average maximum intensity of the peak 'B' as a dose function for all detectors' types.

It is worth mentioning also that at doses of a few kGy gradual yellowing of the detectors have been observed, as the dose increases the colour darkens, reaching dark brown for doses close to1 MGy. This effect was observed for all detectors types. However, the shade of colour depends on the concentration of dopants also as this phenomenon relies on the colour centres formation under the influence of ionizing radiation. These centres alone or in agglomerations with other defects constitute centres of luminescence in TL materials [Schulman and Compton, 1962; Townsend and Kelly, 1973; McLaughlin, 1996]. However, after being heated up to 550ºC during readout all detectors regained their original colour. This is probably caused by the annihilation of the F centres with the H centres, which removes lattice defects [McKeever, 1985]. This effect is under studies recently [see also Remy et al., 2016].

## 4. Summary

Based on the obtained results it has been found that in case of high doses both magnesium and copper are essential for high-temperature TL signal. Analysis of the glow-curves of detectors with an altered concentration of magnesium led to the conclusion that magnesium largely affects the formation of high-dose high temperature peaks. The observed total TL signal increases with increasing dose for detectors with reduced concentration of magnesium. Initially for doses up to 30 kGy, the signal is lower than for the reference detectors. This is coming in line with smaller sensitivity of these detectors to small doses. Very good results are obtained for the highest doses, where the glow-curve with a well distinguished peak "B" of high intensity is observed. For detectors with increased concentration of Mg there is no clear relationship. It can be noted that the highest total TL signal was obtained for a few kilogray while the smallest for hundreds kilogray range.

Lowering the concentration of copper in the detector causes that the resultant TL signal is higher than for the reference detector for doses lower than 5 kGy. Its value, however, decreases with increasing dose reaching the lowest value (approx. 80% in relation to the reference detector) for the highest dose studied. An inverse relationship was obtained in case of increased copper concentration in detector. However, the values of TL signal integrals and the peak 'B' intensity obtained still achieve higher values than for the reference detector. These detectors also showed a good value of the sensitivity to low doses.

Concerning the TL peak 'B' in the case of the reference detectors, 0C and 0.5C detectors, it is observable already at a dose of 50 kGy, whereas for detectors type 2C, 2M and 0.5M this happens only at a dose of 100 kGy. With increasing dose shift of the peak 'B' position to higher temperatures occur for each type of detector. Peak position shift towards higher temperatures with respect to the reference detector is visible for 0C and 0.5C detectors type, while towards the lower for 2M. In other cases, this shift is not clear. The highest intensity of peak 'B' was obtained at a reduced concentration of magnesium (0.5M). For detectors with altered concentration of copper the best values were obtained with increased concentrations of this dopant.

Major changes in the glow-curves of detectors with altered concentrations of magnesium testify to the fact that magnesium is the dominant dopant of LiF based phosphors for high doses. Particular impact is visible in the peak 'B', it has received the highest intensity for the detectors with reduced concentration of magnesium. It should be noted, however, that for this detector TL signal deteriorates at lower doses (its' improvement was seen with the increased concentration of magnesium). It can therefore be concluded that the concentration of magnesium significantly affects the high-dose high temperature peaks. In contrast, in the case of copper, the increase in concentration results in improved results over the whole range of doses. The resulting change for the highest doses are not as significant as for the reduction of Mg.

We determined the effect of concentration of specific dopants on the TL peak 'B'. It was found that both magnesium and copper are necessary to obtain high-dose TL signal. With the reduced concentration of copper the shape of the glow-curve does not differ significantly from the standard one for high doses. In addition for slightly reduced copper concentration the TL peak ‚B' is better separated from the rest of the glow-curve. For an elevated concentration of Cu the TL peak ‚B' is shifted towards lower temperatures in comparison to reference detectors. This difference decreases with increasing dose. Increased concentration of Mg causes a large loss of high-temperature signal and the difficulty of separating the peaks of the glow-curve. However, the important for high-dose TL features seems to be also concentration of phosphorus, but these studies will be published separately, the same concern results of TL emission spectral studies for all presented detectors batches. The important issue is also a detailed examination of the high-dose TL signal while simultaneously changing the concentration of the two dopants which is in progress.

**Acknowledgments**

This work was partly supported by a research project from the National Science Centre, Poland (Contract No. 2013/09/D/ST2/03718).